\input{psfig.sty}
\documentclass[12pt,preprint]{aastex}
\usepackage{color}

\shorttitle{X-Ray Observations of Cygnus~X-2}
\shortauthors{Piraino et al.}

\begin{document}

\title{X-Ray Spectral and Timing Observations of Cygnus~X-2}

\author{S. Piraino\altaffilmark{1,2,3}, A.
Santangelo\altaffilmark{2}, P. Kaaret\altaffilmark{1,4}}

\altaffiltext{1}{Harvard-Smithsonian Center for Astrophysics,
60 Garden St., Cambridge, MA 02138, USA}

\altaffiltext{2}{Istituto di Fisica Cosmica con Applicazioni
all'Informatica - Consiglio Nazionale delle Ricerche, Via Ugo
La Malfa 153, 90146 Palermo, Italy}

\altaffiltext{3}{e-mail: piraino@ifcai.pa.cnr.it}

\altaffiltext{4}{e-mail: pkaaret@cfa.harvard.edu}

\begin{abstract}

We report on a joint BeppoSAX/RossiXTE observation of the
Z-type low mass X-ray binary Cyg~X--2. The source was in the
so-called high overall intensity state and in less than 24
hours went through all three branches of the Z--track.  The
continuum x-ray spectrum could be described by the absorbed
sum of a soft thermal component, modeled as either a
blackbody or a multicolor disk blackbody, and a Comptonized
component. The  timing power spectrum showed several
components including  quasiperiodic oscillations in the range
28--50~Hz while the source was on the horizontal branch
(horizontal branch oscillation; HBO).   We found that the HBO
frequency was well correlated with the parameters of the soft
thermal component in the x-ray spectrum.  We discuss
implications of this correlation for models of the HBO.

\end{abstract}


\section{Introduction}

Cygnus X--2 is a bright persistent Low--Mass X--ray Binary
(LMXB). It consists of a neutron star, having the highest
optically measured mass reported to date ($M_X>1.88$
$M_\odot$, Casares, Charles \& Kuulkers 1998), orbiting the
evolved, late--type companion V1341 Cyg, with a period of
$\sim 9.8$ days (Cowley, Crampton \& Hutchings 1979; Casares,
Charles \& Kuulkers 1998). Although Cyg X--2 is one of
longest know and best studied LMXBs, a clear understanding of
its behavior has not yet been obtained.

Cyg X--2 is classified as a Z-source due to its behavior when
studied on an x-ray color-color diagram (CCD).  Z-sources
follow a one-dimensional trajectory in the two-dimensional
CCD, tracing out a Z shape as the source evolves between
different spectral states.  The parts of the Z are referred
to as the horizontal branch (HB) at the top of the Z, the
normal branch (NB) along the diagonal of the Z, and the
flaring branch (FB) at the base of the Z.  Z-sources move
continuously along the Z-diagram and the position of the
source within the Z is believed, from simultaneous multi-band
(radio, UV, optical and X--ray) observations (Hasinger et al.
1990), to be related to the mass accretion rate with
accretion at a minimum at the left end of the HB and at a
maximum at the right end of the FB.   However, the behavior
of Z-sources is not described by a single parameter because,
on time scale of weeks to months, the morphology and position
of the Z-track in the CCD vary.  Kuulkers et al.\ (1996)
described the phenomenology for Cyg X--2 in terms of three
overall intensity levels (high, medium, and low), with
different morphology for the Z track in the  different
intensity levels.

Z-sources, including Cyg X--2, show complex X-ray spectrum,
requiring a superposition of several spectral components.
Different composite models, including thermal bremsstrahlung
together with blackbody radiation, two blackbodies, a power
law plus thermal bremsstrahlung, a multi--temperature
blackbody disk plus boundary layer blackbody, or a
Comptonized spectrum, have all been found to describe
adequately the data.  Unfortunately, spectral modeling alone
has not led to unique identification of the physical origin
of the various spectral components.  Further observations
relating the x-ray spectra to other other source parameters
are required.

Z-sources also show rich X-ray timing power spectra including
several components and varying with position in the CCD. Cyg
X--2 shows quasiperiodic oscillations (QPOs) with frequency
varying between $\sim 15$ Hz and $\sim 60$ Hz, present on the
horizontal branch and on the upper part of the normal branch
(HBOs), and high frequency QPOs (kHz QPOs), ranging between
300 Hz and 900 Hz have been detected on the HB (Wijnands et
al.\ 1998).  Slower QPOs with frequencies of 5--7 Hz are
observed on the lower part of the normal branch (called
NBOs). Sometime HBOs are observed simultaneously with the
NBOs, indicating these QPOs should be produced by different
mechanisms.  Agreement is lacking on a unique model to
explain the HBOs, the NBOs, or the kHz QPOs.  Correlations
between the timing parameters and other source parameters may
help limit the number of viable models for the QPOs.

In this paper, we present joint timing and spectral
observations of Cygnus X--1 obtained with RossiXTE and
BeppoSAX.  The goal of our observations was to obtain
simultaneous x-ray spectral and timing data for Cyg X-2 in
order to study correlations between the spectral and timing
parameters.  The observations and analysis are described in
\S 2.  Results on the color and intensity evolution are
presented in \S 3, on timing power spectra in \S 4, and on
energy spectra in \S 5.  The correlations between spectral
and timing properties is described in \S 6. We discuss the
implications of our results in \S 7.

\section{Observations and Analysis}

On 2000 June 2--3, we performed a joint observation of Cyg
X--2 with the RossiXTE and BeppoSAX satellites. The pointing
was carried out between 2000 June 2 13:11 UT and June 3 17:23
UT, with RossiXTE (Bradt et al. 1993) and between 2000 June 2
18:56 UT and June 3 14:25 UT with the BeppoSAX Narrow Field
Instruments, NFIs (Boella et~al.\ 1997).

Spectral data obtained from the RXTE Proportional Counter
Array (PCA) in 129 photon energy channels with a time
resolution of 16 s (standard 2) were used to extract a
color-color diagram (CCD) and a hardness-intensity diagram
(HID). The soft and hard colors were defined as counts rate
ratios between 4.0--6.4~keV and 2.6--4.0~keV and between
9.7--16~keV and 6.4--9.7~keV. The intensity was defined as
the PCU2 count rates in the energy band 2.6--16~keV. All
count rates were background subtracted.  We used only PCU2
for the intensity and color analysis as only it and PCU0 were
on during all observations and PCU0 suffered from a propane
layer leak which made the energy calibration very uncertain.

Timing data obtained from the PCA with a time resolution of
125~$\mu s$ in Single Bit (one energy channel) modes for the
energy bands 2.0--5.8~keV, 5.8--10.1~keV, 10.1--21.4~keV, and
21.4--120~keV were used to obtain power spectra.  We note
that the data mode used does not allow us to remove events
from PCU0.  Thus, due to the propane layer leak, the true
energy bands for PCU0 likely differ from these nominal
ranges.  However, our basic result on the energy dependent of
the timing noise (that the noise features are stronger at
higher energies) is not sensitive to the details of the
energy band boundaries.

To search for fast kHz QPOs, we performed Fast Fourier
transforms on 2~s segments, added all power spectra obtained
within each selected observation interval, and then searched
for excess power in the 200--2048~Hz range. No kHz QPOs with
a chance probability of occurrence of less than $1\%$ were
found either in the full band data or in the data above
5.8~keV. The kHz QPO previously found from Cyg X--2 occurred
in the medium overall intensity state (Wijnands et al.\
1998).  No kHz QPOs have been observed during the high
intensity state.

To study the low frequency behavior, we performed Fast
Fourier transforms on 64~s segments and we added the 64~s
power spectra obtained for all or the three highest energy
bands within each continuous observation interval. The
expected Poisson noise level, due to counting statistics, was
estimated taking into account the effects of the deadtime and
subtracted from each power spectrum, and the power spectra
were normalized to fractional rms.  Each 0.1--100~Hz total
power spectrum was fitted as described below.

Data from the BeppoSAX NFIs were used to extract broad band
energy spectra. The NFIs are four coaligned instruments
covering the 0.1--200~keV energy range: LECS (0.1--10~keV),
MECS (1.3--10~keV), HPGSPC (7--60~keV), and PDS
(13--200~keV). The net on--source exposure times were 11~ks,
27~ks, 35~ks and 20~ks for LECS, MECS, HPGSPC, and PDS
respectively. LECS and MECS data were extracted in circular
regions centered on Cyg X--2's position having radii of $8'$
and $4'$ respectively.  The same circular regions in blank
field data were used for the background subtraction.  Cyg
X--2 is sufficiently bright during this observation that the
background subtraction does not affect the fits results.
Spectra accumulated from Dark Earth data and during
off--source intervals were used for the background
subtraction in the HPGSPC and in the PDS respectively. The
LECS and MECS spectra have been rebinned to sample the
instrument resolution with the same number of channels at all
energies.  Logarithmic rebinning was used for the HPGSPC and
PDS spectra.

We investigated the source power and energy spectrum in
different regions of the HID. Energy spectra of each NFI were
create over five different HID intervals where BeppoSAX data
were present. These spectra are indicated with E, F, C, B, A and
reported in order of position along the Z track in Tables 3 and
4. The alphabetical order indicates the sequence in time (see
MECS light curve in Figure 2).

\section{Color and Intensity Evolution}

Figure 1 shows the RXTE ASM light curve of Cygnus X--2 where
the long--term X-ray variation is clearly visible.  The time
of the RXTE/BeppoSAX observation is indicated with an arrow
on the right.  The light curve around the observation (see
the inset in Figure 1) shows complex time variability. The
source was in the high overall intensity state (Kuulkers, van
der Klis \& Vaughan 1996) and during the observation moved
along the three branches of the Z pattern (this is more
evident in the HID than in the CCD).  No x-ray bursts
occurred during the observation.

The RXTE/BeppoSAX light curves obtained in the entire energy
range of the PCU2 and the various energy ranges of the NFIs are
shown in Figure 2, and Figure 3 shows the color-color (CCD) and
hardness-intensity (HID) diagrams. The bin size is 256 s. In
Figure 3, each symbol indicates a different continuous segment
in the light curve (see upper panel).

The states of Z sources are typically described in terms of a
horizontal branch (HB), a normal branch (NB), and a flaring
branch (FB), as described above. The HB consists of the
points with high hardness and varying intensity in the HID,
the NB has high intensity and varying hardness, and the FB
has low hardness and varying intensity. As observed during
other high intensity levels (Wijnands et al.\ 97, Wijnands \&
van der Klis 2001) the FB in HID is almost horizontal and the
correlation of intensity with position along the Z track
reverses in the FB relative to the HB.  We divided the data
into seven states for use in the analysis below.  Four
regions are in the HB (marked as {\tt hb1, hb2, hb3, hb4} on
the HID) one is in the NB ({\tt nb1}), one region is at the
NB/FB corner ({\tt nbfb}), and finally one is in the FB ({\tt
fb1}).

%The FB is blended with the NB in the CCD, while it is
%extended and clearly visible in the HID. The FB in HID is
%almost horizontal and the intensity is anticorrelated with
%the mass accretion rate.

The overall intensity of the source changes by more than a
factor of 2 during the entire observation (see the PCU2 light
curve). The character of the intensity variations changes with
energy. Just after the beginning of the BeppoSAX observation,
near a time of $2 \times 10^4 \rm \, s$ and marked as {\tt nbfb}
in Figure 2, the source moved from the flaring branch (FB) to
the normal branch (NB) and the counts rates increased by more
than a factors of 5 in the 4--10~keV, 10--15~keV, 15--50~keV
energy ranges but remained roughly constant in the 0.2--4~keV
band. Conversely, near a time of $5 \times 10^4 \rm \, s$, and
marked as {\tt hbnb} in Figure 2, the source crossed the NB/HB
corner and the count rates in the bands below 10~keV suffer
large decreases, while the rate in the 15--50~keV band remains
decreases only slightly.

\section{Fast Timing}

To investigate the evolution of fast timing behavior with
position in the HID, we created power spectra in 7 regions of
the HID described above and shown in Figure 3. The resulting
power spectra obtained from the full PCA band are shown in
Figure 4 and power spectra obtained using events from the three
higher energy single-bit channels, i.e. photons with energies
above 5.8~keV, in Figure~5.

At least five different power spectral components have been
distinguished in the power spectra of Cygnus X--2 and other Z
sources (Hasinger and van der Klis 1989): 1) a power law
which is referred to as Very Low Frequency Noise (VLFN)
because it dominates below 0.1~Hz; 2) a broad noise
component, referred as Low Frequency Noise (LFN), represented
by a flat power law with a high frequency cutoff, which
characterizes the HB power spectra up to around 10~Hz; 3) a
broad noise component having a similar shape as LFN, but
extending to 100~Hz or above, known as High Frequency Noise
(HFN), present in all spectral states; 4) a narrow excess
with centroid frequency in the 15--60~Hz range, present in
the HB, called the horizontal branch oscillation (HBO), and
modeled with a Lorentzian; 5) another broad peak present in
the NB, similar in shape to HBO, but with a characteristic
frequency of around 6~Hz, called the normal branch
oscillation (NBO). The solid curves in Figures 5 and 6
represent analytic fits consisting of these components. The
components used for each power spectrum and the fit
parameters obtained are presented in Tables 1 and 2. The
fractional rms amplitude for the VLFN component was found by
integrating over the frequency range 0.001--1~Hz. A range of
0.01--100~Hz was used for both LFN and HFN components.
Because the HFN powerlaw index $\alpha_H$ been found to be
consistent with zero in Z sources (Hasinger \& van der Klis
1989), we fixed it to zero.

A strong narrow HBO is present at the left end of the HB
({\tt hb1}). A faint and narrow feature was visible in the
first three HB power spectra near the position of the second
harmonic, and there is also a possible third harmonic in the
{\tt hb1} power spectrum. Moving along the HB towards the NB,
the frequency increased from 28~Hz to 50~Hz, while its rms
amplitude decreased from 2.8~$\%$ to 1.7~$\%$ (for $E >
5.8$~keV, the rms decreased from 4.3~$\%$ to 2.4~$\%$), and
its width increased from 6.7~Hz to 16.5~Hz. This behavior is
consistent with that observed in the Ginga data by Wijnands
et al.\ (1997).

LFN and HFN can be identified in all 4 HB power spectra. The
VLFN was not fitted separately from LFN because its power in
the 0.1--100 Hz range was much smaller than that of LFN. The
rms of the LFN was 4--5\% (6--7\% for $E > 5.8$~keV) and
changed little along the HB. The powerlaw index
$\alpha_{LFN}$ and the cut--off frequency $\nu_{cLFN}$ showed
a positive correlation with the HBO frequency. The HFN
component showed a cutoff frequency $\nu_{cLFN}$ which is
consistent with constant within the (sometimes large) errors,
and a clear inverse correlation between its strength and the
HBO frequency.

On the upper part of the NB ({\tt nb1}), the HBO is still
visible at 54 Hz with an rms level of 1.8\% (3.1\% for $E >
5.8$~kev). In addition, a broad NBO is present at $\sim$5~Hz.
The NBO feature is lost in the continuum as the source moved
towards the FB. A very strong VLFN component (rms up to $\sim
17.4 \%$) replaced the LFN in the continuum of the NB and FB
power spectra. The VLFN became stronger moving from the NB to
the FB, and the HFN showed the opposite trend.

\section{Spectral Results}

To investigate the spectral behavior versus the position on the
Z track, we created energy spectra in 5 regions of the HID
described above. We fitted the X-ray continuum of Cyg X--2 with
several two--component models. A blackbody or disk multicolor
blackbody ({\tt diskbb}, Mitsuda et al. 1984) was used to model
the soft component, and one of three thermal Comptonization
models, a power law with cutoff, or the solution of the
Kompaneets equation given by Sunyaev \& Titarchuk 1980 {\tt
compst} or its improved version {\tt comptt} (in which
relativistic effects are included, Titarchuk 1994), was used for
the harder component. The addition of a gaussian around 1~keV
was required in all intervals. Lines near 1~keV have been
observed previously from Cyg X-2 with instruments with higher
resolution (Vrtilek et al.\ 1988). We also included a gaussian
near 6.7~keV to model possible Fe emission lines. The best fit
to the continuum of Cyg X--2 was obtained using a blackbody or a
diskbb plus the {\tt comptt} Comptonization model (thereafter we
call {\tt bb+comptt} and {\tt diskbb+comptt}).

Using the {\tt diskbb+comptt} model (Table~4, plots on the right
in Figure~7) adequate fits were obtained. There is a slight, but
not statistically significant excess of counts at high energy in
the E and F spectra. The parameters of the {\tt diskbb}
component varied significantly with position along the Z track
with the inner disk temperature, $kT_{in}$, changing from
0.79~keV to 1.07~keV, and the inner radius $R_{in}$ changing
from 63~km to 47~km. Strong correlations of the parameters of
the {\tt comptt} component, other than its flux, with position
along the Z track were not evident. The temperature and
effective Wien radius of the seed photons for the Comptonized
component were constant at $kT_{\rm W}\sim 1.2$~keV and $R_{\rm
W} \sim 23~km$. The temperature of the Comptonizing region was
also roughly constant at $k T_{\rm e} \sim 3$~keV, while the
optical depth varied in the range $\tau \sim 7-10$.

With the {\tt bb+comptt} model, an excess of counts above
$\sim 30$~keV was apparent in all five spectra. Because it is
now known that a hard tail is sometimes present in the
spectra of Z sources, we added a powerlaw component to fit
the hard tail. Addition of the powerlaw is justified by the
F-test statistic (the probability that the improvement in
$\chi^2$ is due to chance is always less than $3 \times
10^{-5}$). This additional component was detected up to
energies of $\sim 80-100$~keV, had a best fit photon index
$\sim 2.7$ and contributed $\sim 14\%$ of the 0.1-100~keV
source flux.

Hard tails have been reported in the spectra of Z sources,
including for GX~5-1 using Ginga by (Asai et al.\ 1994), for
Cyg X-2 using BeppoSAX (Frontera et~al.\ 1998) , for Sco X-1
using CGRO/OSSE (Strickman \& Barret 2000) and using
RXTE/HEXTE (D'Amico et al.\ 2001a), and for GX 17+2 (Di Salvo
et al.\ 2000), GX 349+2 (Di Salvo et al.\ 2001), and Cir X-1
(Iaria et al.\ 2001) using BeppoSax observations. The need
for a hard powerlaw tail in the present data depend on the
choice of the spectral model, and, thus, cannot be considered
a detection of the hard powerlaw tail. D'Amico et al.
(2001b), taking into account all the results on the
detections of hard X-ray tails in Sco X-1 and GX 349+2, argue
that the appearance of a such component is correlated with
the brightness of the thermal component. Specifically, they
propose that the production of a hard X-ray tail in a Z
source is a process triggered when the thermal component is
brighter than a level of $\sim$ 4$\times$ 10$^{36}$ ergs
s$^{-1}$ in the 20--50 keV energy range. In this energy
range, the luminosity of Cyg X-2, in the observation
discussed here, was $\lesssim$ 4.5 $\times$ 10$^{36}$ ergs
s$^{-1}$, i.e.\ very close to the trigger value. Hence, our
results appear consistent with previous observations.
However, the fact that the hard powerlaw component is not
required in the {\tt diskbb+comptt} model underscores the
importance of understanding the low energy emission in
modeling the hard x-ray emission from x-ray binaries.

Results from these three-component fits to the BeppoSAX
spectra of Cyg X--2 are given in Table~3 and in the plots on
the left in Figure~7. Because the powerlaw normalization,
absorption column $N_H$, and Luminosity, obtained by fitting
the E segment energy spectrum, were surprisingly higher than
in other segments, we also report the fit parameters obtained
by fixing the $N_H$ value to 2$\times10^{21}$ cm$^{-2}$,
close to the average value from the other observations.  The
blackbody flux is well correlated with position along the Z
track. The blackbody temperature and equivalent radius were
approximately constant at $kT_{\rm BB}\sim 0.5$~keV and
$R_{\rm BB}\sim 70$~km (using a distance of 11.6 kpc, Casares
et al. 1999), respectively. The temperature of the seed
photons for the Comptonized component was roughly constant at
$kT_{\rm W}\sim 1$~keV. Following In 't Zand et al.\ (1999),
we derived an effective Wien radius of the seed photons; this
was roughly constant at $R_{\rm W} \sim 30$~km. The
temperature and optical depth of the (spherical) Comptonizing
region were $k T_{\rm e} \sim 3$~keV and $\tau \sim 9$,
respectively.

The addition of an Fe K$_\alpha$ line at $\sim 6.7$~keV
proved necessary, especially when the source was in the HB
(spectra E and F). Similar parameters for the line where
found using both the {\tt bb+comptt} and {\tt diskbb+compTT}
models. BeppoSAX does not have adequate spectral resolution
to draw detailed conclusions about the origin of this line
(Piraino, Santagelo, \& Kaaret 2000), but the presence of the
line is motivation for high resolution spectral studies of
Cyg X-2 while in the HB.

\section{Correlation of Spectral and Timing Parameters}

To directly compare the spectral and timing properties of Cyg
X-2, we performed a second timing analysis of the RXTE data
using regions of the Z-diagram matching the regions selected
for spectral analysis. The analysis was performed as
described above. In Figure 8, we plot the spectral parameters
obtained for the Comptonization and thermal components of the
E, F, C, and B spectra for the two models discussed above
versus the HBO frequency.  The flux from the Compton
component appears correlated with HBO frequency for the {\tt
bb+compTT} model, but may not be for the {\tt diskbb+compTT}
model. The other Comptonization parameters do not appear
correlated with HBO frequency in either model.  The total
flux is clearly correlated with HBO frequency in both models.
However, we note that these observations sample only a short
time span and correlations between timing parameters and
total flux are generally not robust over long intervals (Ford
et al.\ 1997). Interestingly, a strong correlation exists
between the parameters of the soft spectral component, either
the blackbody flux for the {\tt bb+compTT} model or the disk
inner radius for the {\tt diskbb+compTT} model, and the HBO
frequency.

\section{Discussion}

We discovered a correlation between the HBO frequency and the
parameters of the soft thermal spectral component -- the
blackbody flux for the blackbody model and the disk inner
radius for the multicolor disk blackbody model. This likely
implies a physical relation between the source of the soft
thermal emission and the timing frequencies.

RXTE and BeppoSAX data from the atoll source 4U~0614+091,
covering a span of more than one year, showed a robust
correlation between the kHz QPO frequency and the flux of the
blackbody component in data spanning several years (Ford et
al.\ 1997; Piraino et~al.\ 1999).  Because the frequencies of
the various timing signals from x-ray binaries appear to be
highly correlated (Psaltis et al.\ 1999), the physical
implications of the spectral-HBO correlation are likely to be
similar to those of the spectral-kHz QPO correlation.

A correlation would occur naturally in models where the
oscillation frequencies are related to the Keplerian orbital
frequency, $\nu_K$, at inner edge of the accretion disk,
$r_{in}$, (e.g.\ Alpar \& Shaham 1985, Miller, Lamb \& Psaltis
1998; Stella \& Vietri 1998) and the flux from the soft thermal
component arises from an accretion disk (the usual
interpretation of the {\tt diskbb} model). The Keplerian
relation, $r_{in} \propto \nu_{HBO}^{-2/3}$, would be expected
if the oscillation frequency, $\nu_{HBO}$, is linearly related
to the Keplerian orbital frequency, $\nu_K$. We fitted a
function of the form $r_{in} = A \nu_{HBO}^{\alpha}$ to the data
with $r_{in}$ taken as the disk radius from the {\tt diskbb}
spectral component. The best fit exponent is $\alpha = -0.50 \pm
0.18$ which is consistent with the Keplerian value $\alpha =
-2/3$. With $\alpha$ fixed to the Keplerian value, the
coefficient $A = 630 \pm 20 \rm \, km \, Hz^{2/3}$. This is
consistent with a $\sim 2 M_{\odot}$ neutron star if $\nu_{HBO}
= 0.2 \nu_{K}$.

In the `magnetospheric beat--frequency model' (e.g. Alpar \&
Shaham 1985; Psaltis et al.\ 1999) the centroid frequency of
the QPO is identified with the difference between the
Keplerian frequency at magnetospheric radius and the spin
frequency, $\nu_S$ of the neutron star. If the inner disk
radius is equal to the magnetospheric radius, then $r_{in} =
(GM/4 \pi^2)^{1/3}(\nu_{HBO} + \nu_S)^{-2/3}$. Fitting this
form to the data, we find $\nu_S = 13 \pm 19 \rm \, Hz$ and
$(GM/4 \pi^2)^{1/3} = 770 \pm 180 \rm \, km \, Hz^{2/3}$.
These values are inconsistent with a neutron star mass above
$0.5 M_{\odot}$.  However, we note that $r_{in}$ may not
reflect the true magnetospheric radius.

The current data consist of only a few points, with
significant errors on the derived spectral parameters for
some of the points. Hence, the data are not sufficient to
adequately constrain the relation between the oscillation
frequency and the spectral parameters so that strong
conclusions about the viability of particular models can be
drawn.  Additional observations with higher spectral
resolution and better low energy response, made
simultaneously with timing observations would be of great
interest and might provide a means to discriminate between
various models of the HBO.

\acknowledgments

We thank Evan Smith and Donatella Ricci for their efforts in
coordinating the observations and the referee for important
comments.  PK and SP acknowledge support from NASA grants
NAG5-9104, NAG5-9097, and NAG5-7405.

\begin{figure}[ht]
\centering \plotone{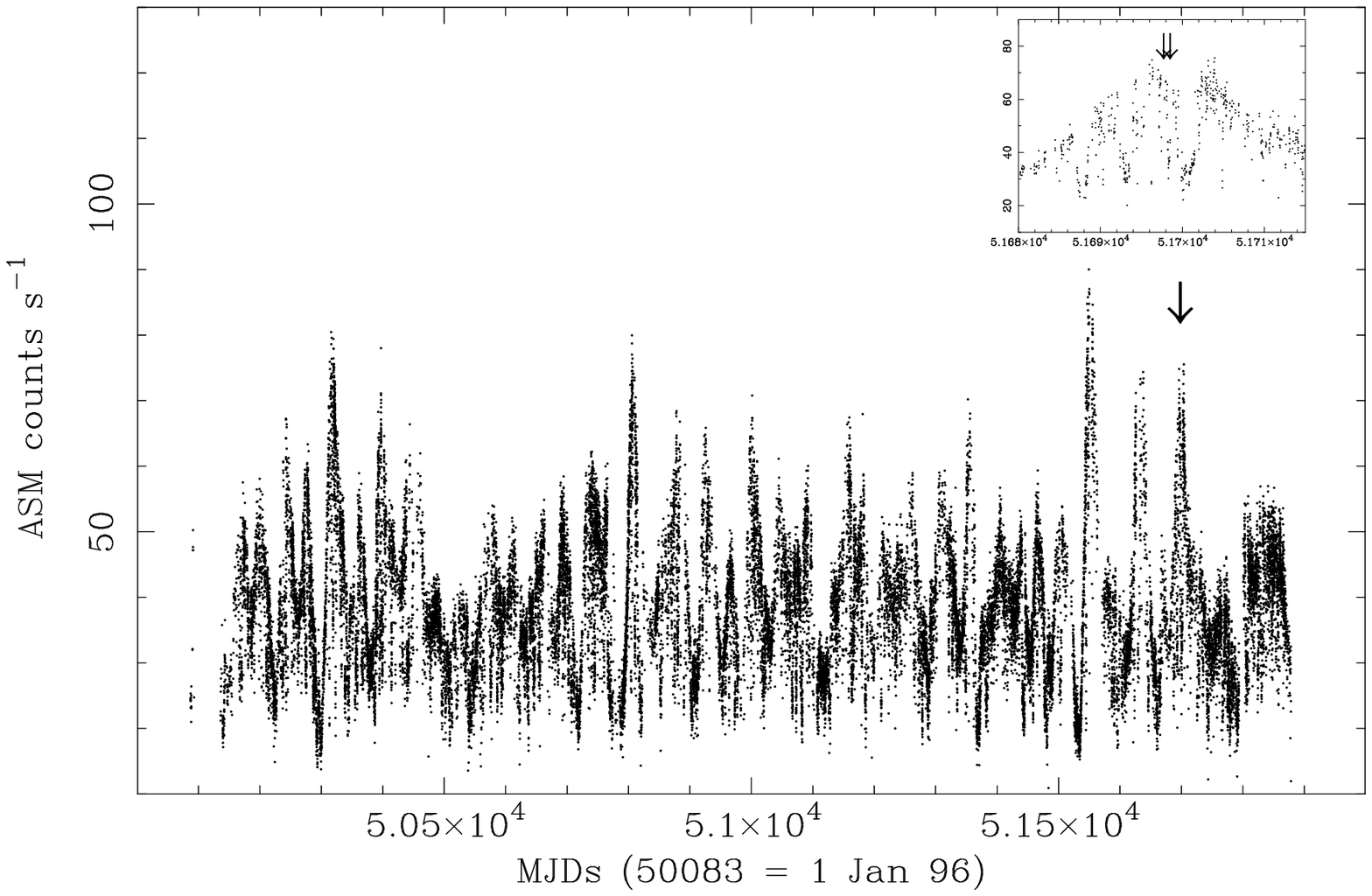}
\caption{RXTE All Sky Monitor
light curve (1.3--12,1 keV) of Cygnus X--2 since January 1996 to
December 2000. Long--term X-ray variations are clearly visible.
The RXTE/BeppoSAX observation is indicated with an arrow on the
right. The inset shows the complex time variability of the
source around the observation.}
\label{fig1}
\end{figure}

\begin{figure}[ht]
\epsscale{0.9} \plotone{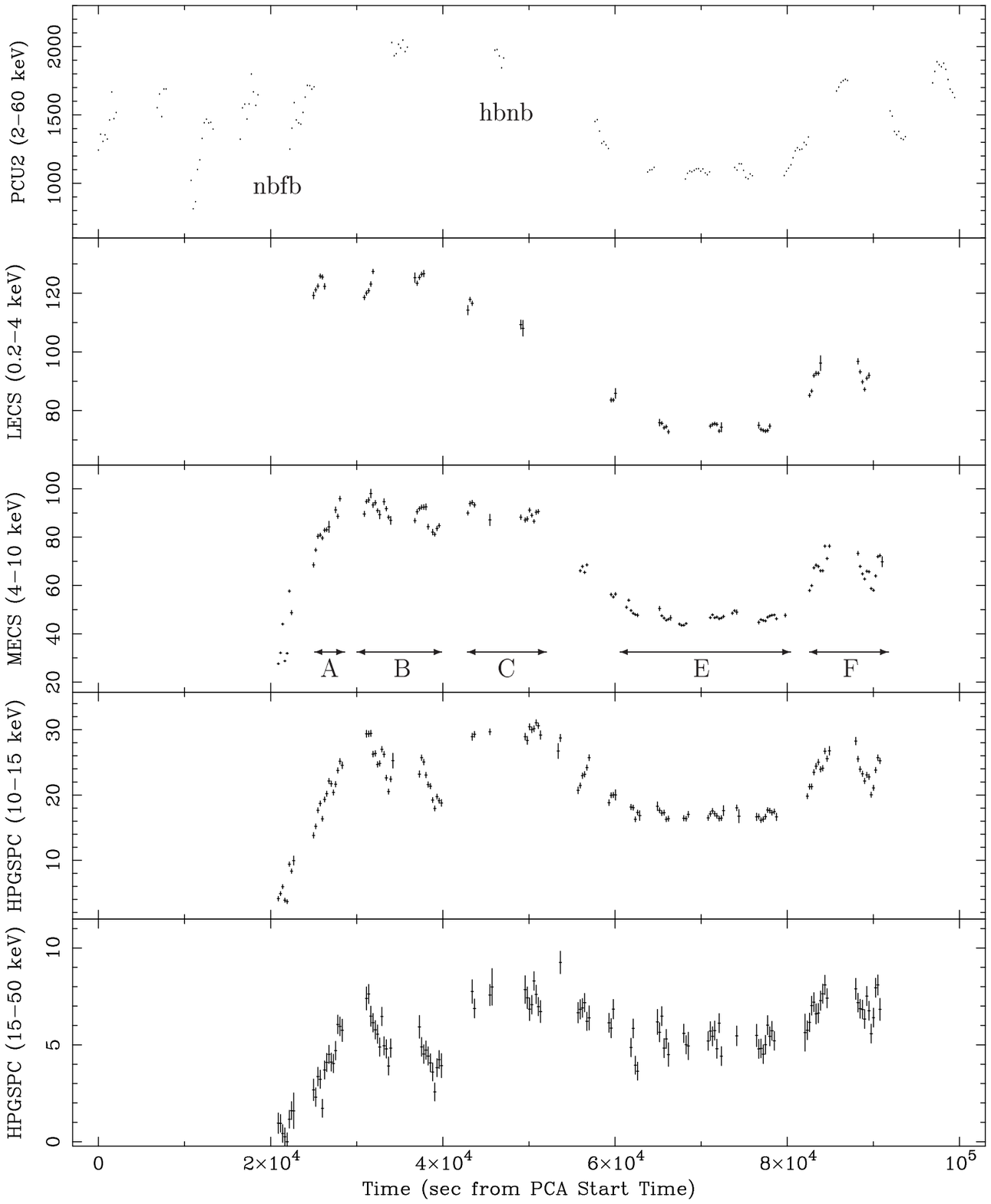} \caption{Light curves of the
Cygnus X-2 in the energy range 2--60 keV of the unit 2 of the
PCA (PCU2, upper panel) and in the four energy ranges 0.2--4
keV, 4--10 keV, 10--15 keV and 15--50 of the BeppoSAX NFIs. Each
bin corresponds to 256 s. The PCA observation starts 20 ks
before than the BeppoSAX observation and ends 10 ks after.}
\label{fig2}
\end{figure}

\begin{figure}[ht]
\centering \plotone{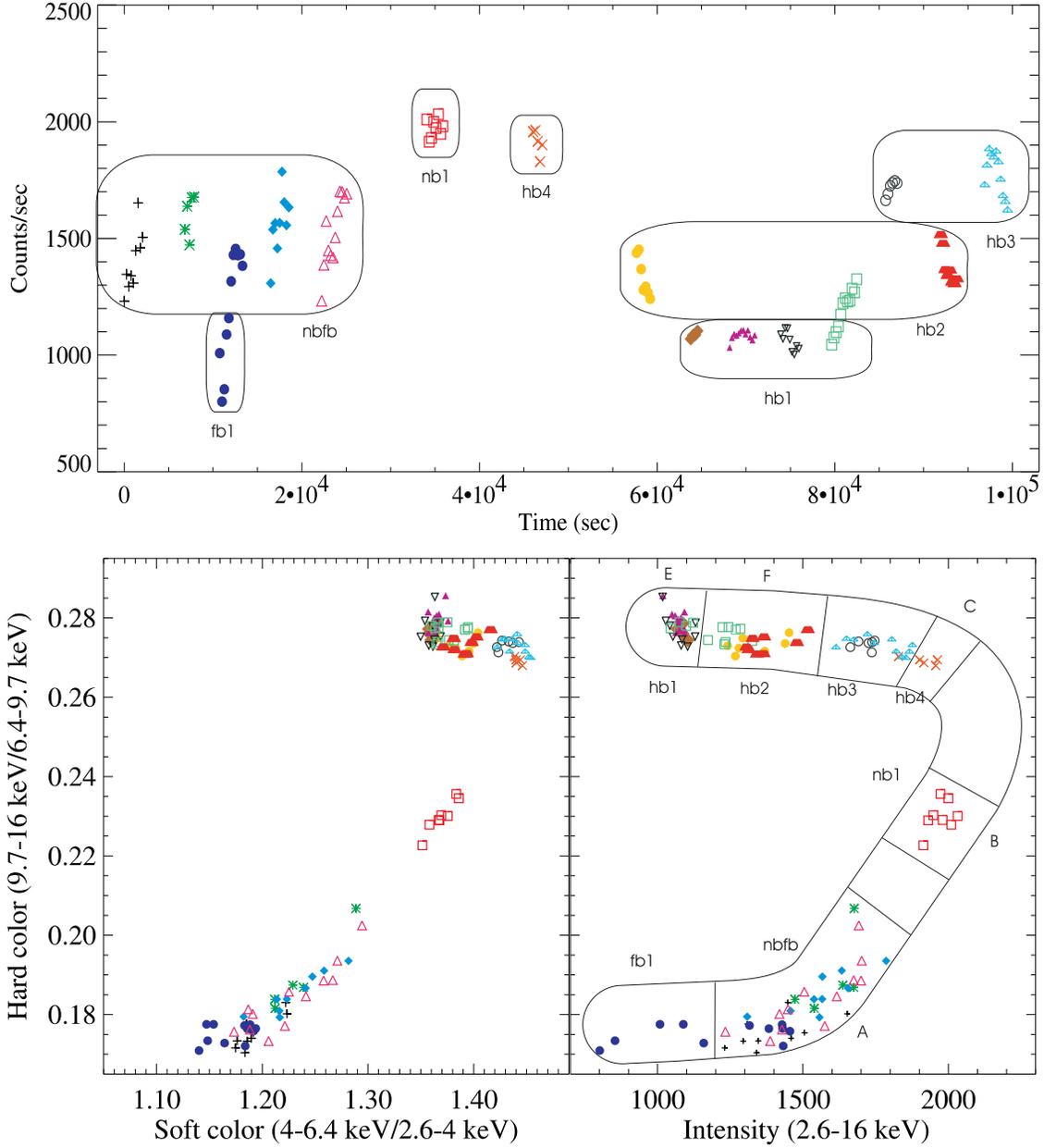} \caption{Light curve (upper panel)
Color--color diagram (bottom left panel) and Hardness--Intensity
diagram (bottom right panel) of Cygnus X--2. The soft color is
the count rate ratio between 4.0--6.4~keV and 2.6--4.0~keV and
the hard color between 9.7--16~keV and 6.4--9.7~keV. The
intensity was defined as the PCU2 count rates in the energy band
2.6--16~keV. All count rates were background subtracted. All
points are 256 s averages. Different colors and markers were
used for different continuous segment in the light curve, as
shown in the upper panel. Data regions used for the timing
analysis are indicated in the HID diagram.} \label{fig3}
\end{figure}
\clearpage

\begin{figure}[ht]
\plotone{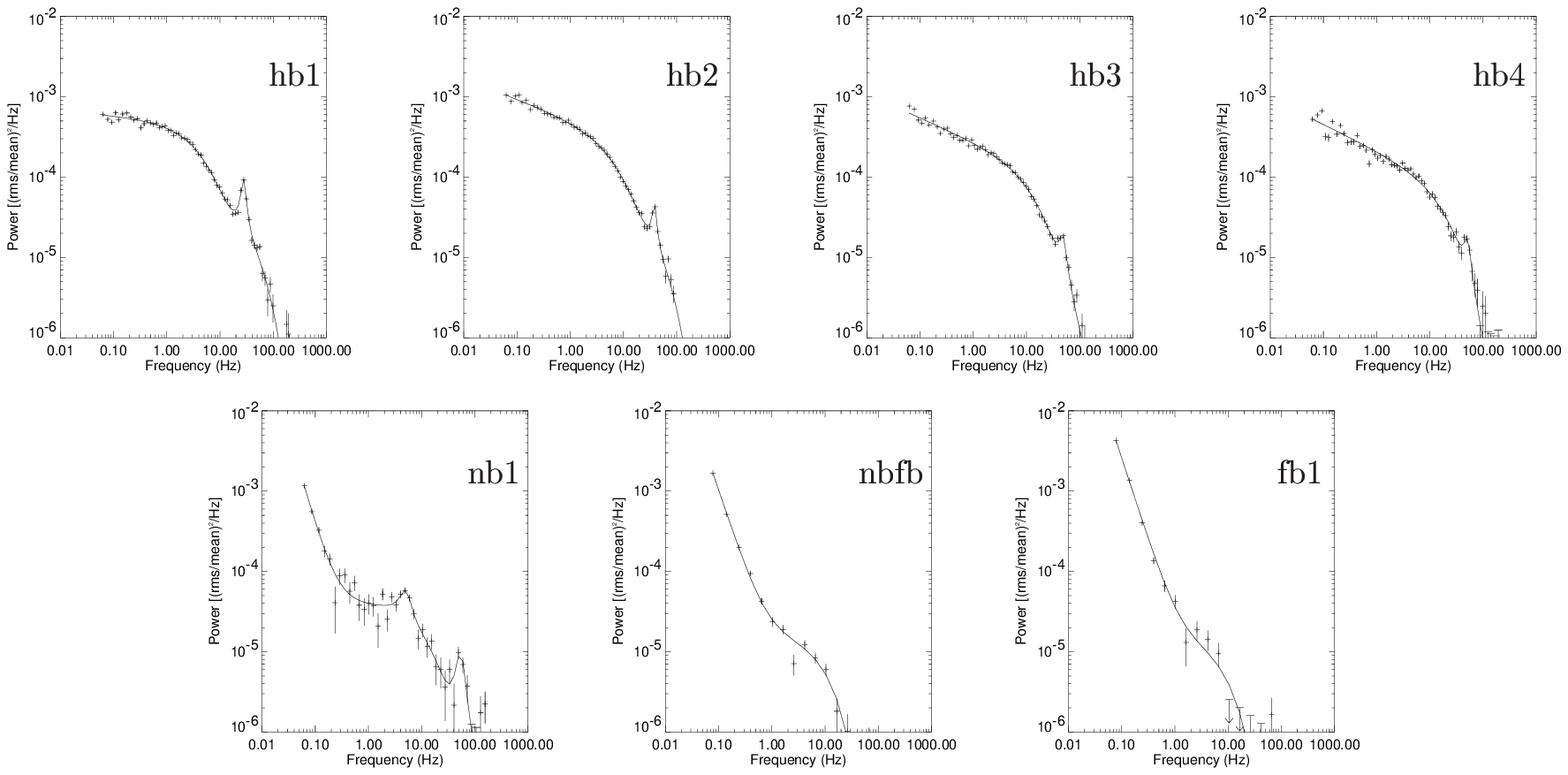}
\caption{Power spectra obtained using data
from all energy bands. The four upper panels show spectra from
the HB, the three lower panels show the NB (left), the NB/FB
(central) and the FB spectra, respectively.} \label{fig4}
\end{figure}
\clearpage

\begin{figure}[ht]
\plotone{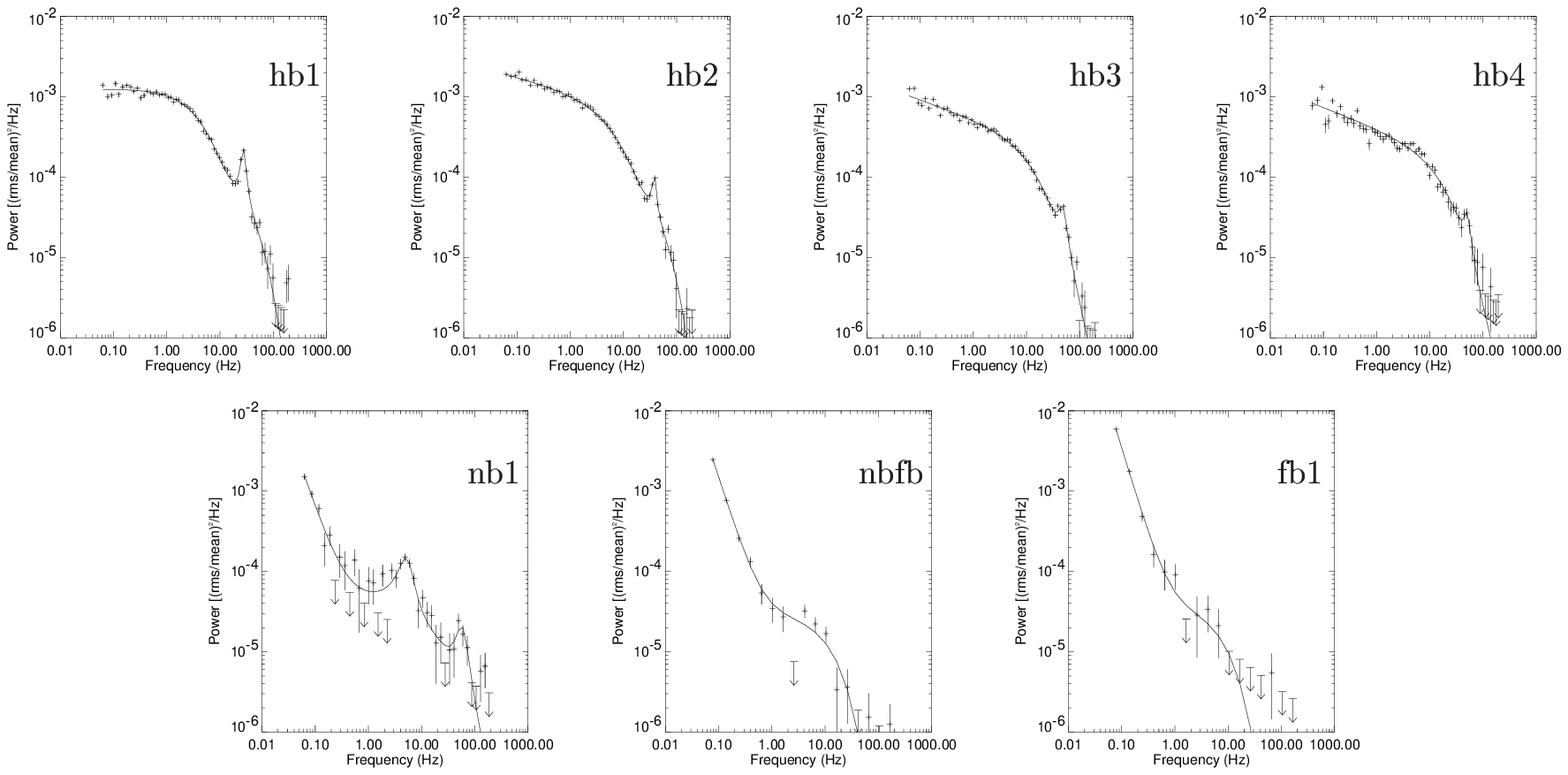}
\caption{Power spectra obtained using
photons with energies above 5.8~keV. The four upper panels show
spectra from the HB, the three lower panels show the NB (left),
the NB/FB (central) and the FB spectra, respectively.}
\label{fig5} \end{figure} \clearpage

\begin{figure}[ht]
\epsscale{0.74}
\plotone{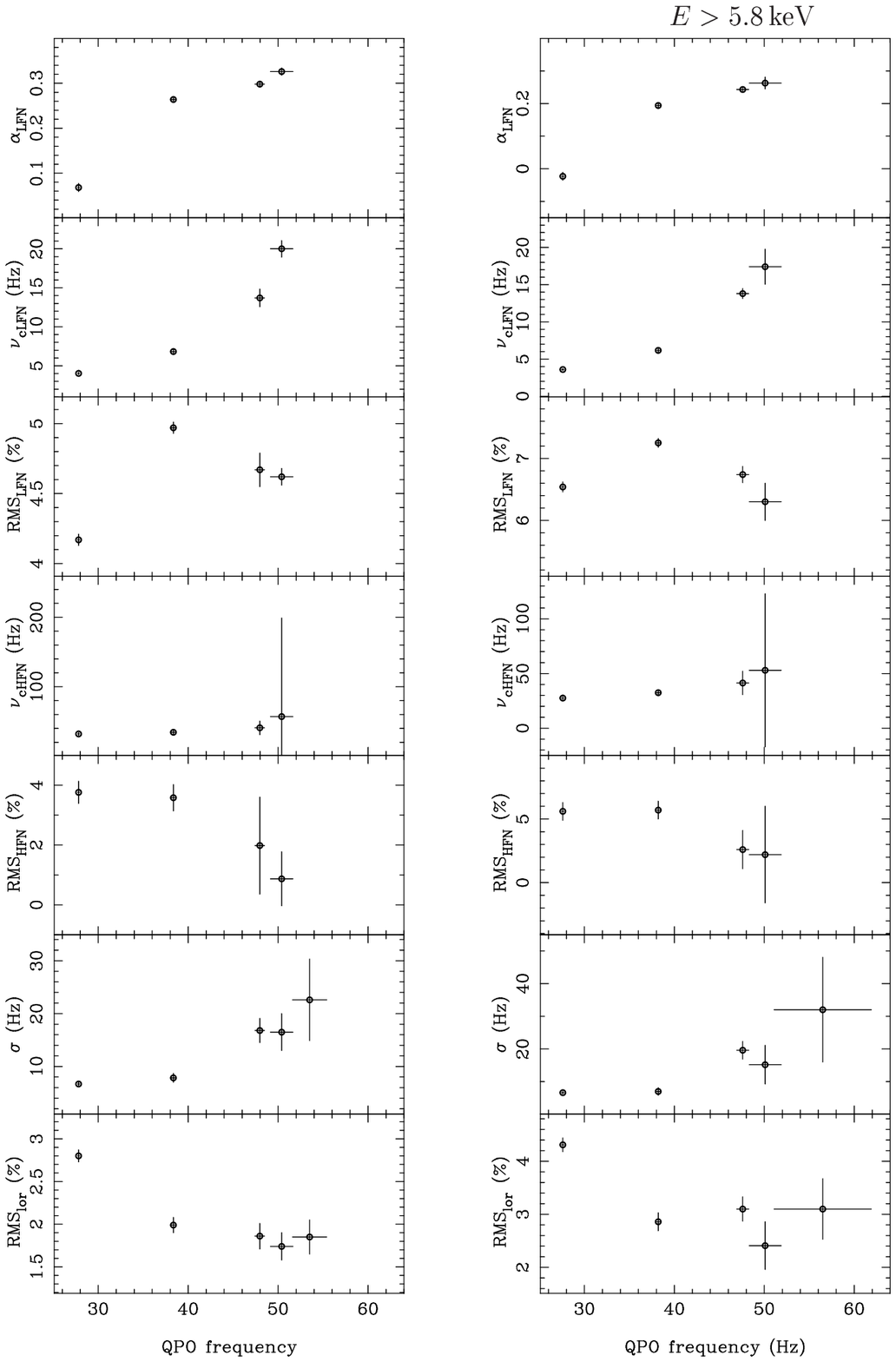} \caption{Best fit
parameters obtained using the standard model for LFN HFN, versus
the HBO frequency. The left panels are relative to the power
spectra extracted using all photons, the right panels are
relative to the power spectra extracted using photons with
energies above 5.8~keV.} \label{fig6} \end{figure}

\begin{figure}[ht]
\epsscale{0.65}
 \plotone{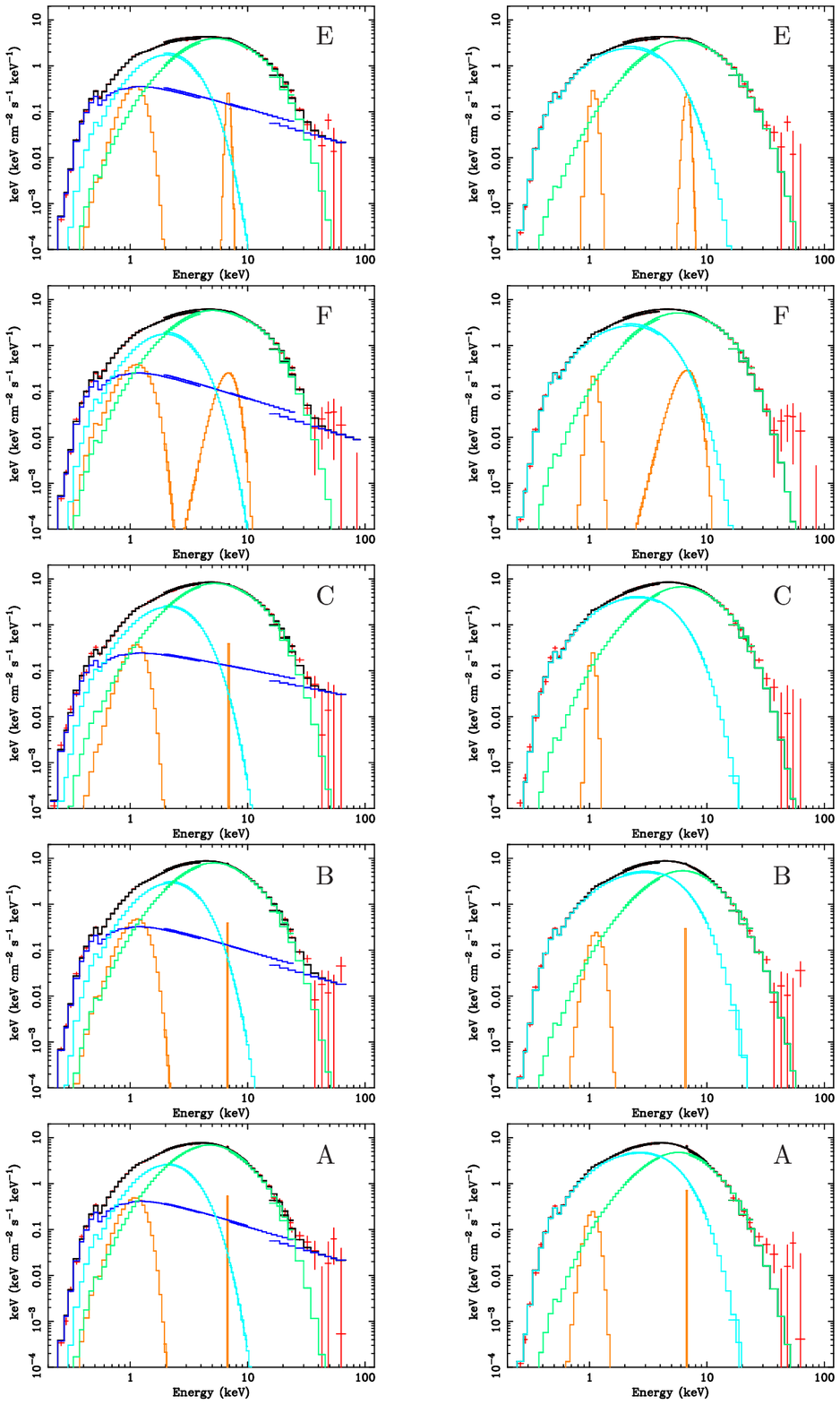}
 \caption{Left: spectra (red color) and
best fit models using \textbf{bb+comptt+po} model.  Right:
spectra and best fit models using \textbf{diskbb+comptt}. The
total model fit is shown using as a black line, the  emission
lines  as orange line, the disk model as light blue line, the
Comptonization component as a green line and  finally the
powerlaw as blue line. } \label{fig7}
\end{figure}

\begin{figure}[ht]
\epsscale{0.75}
 \plotone{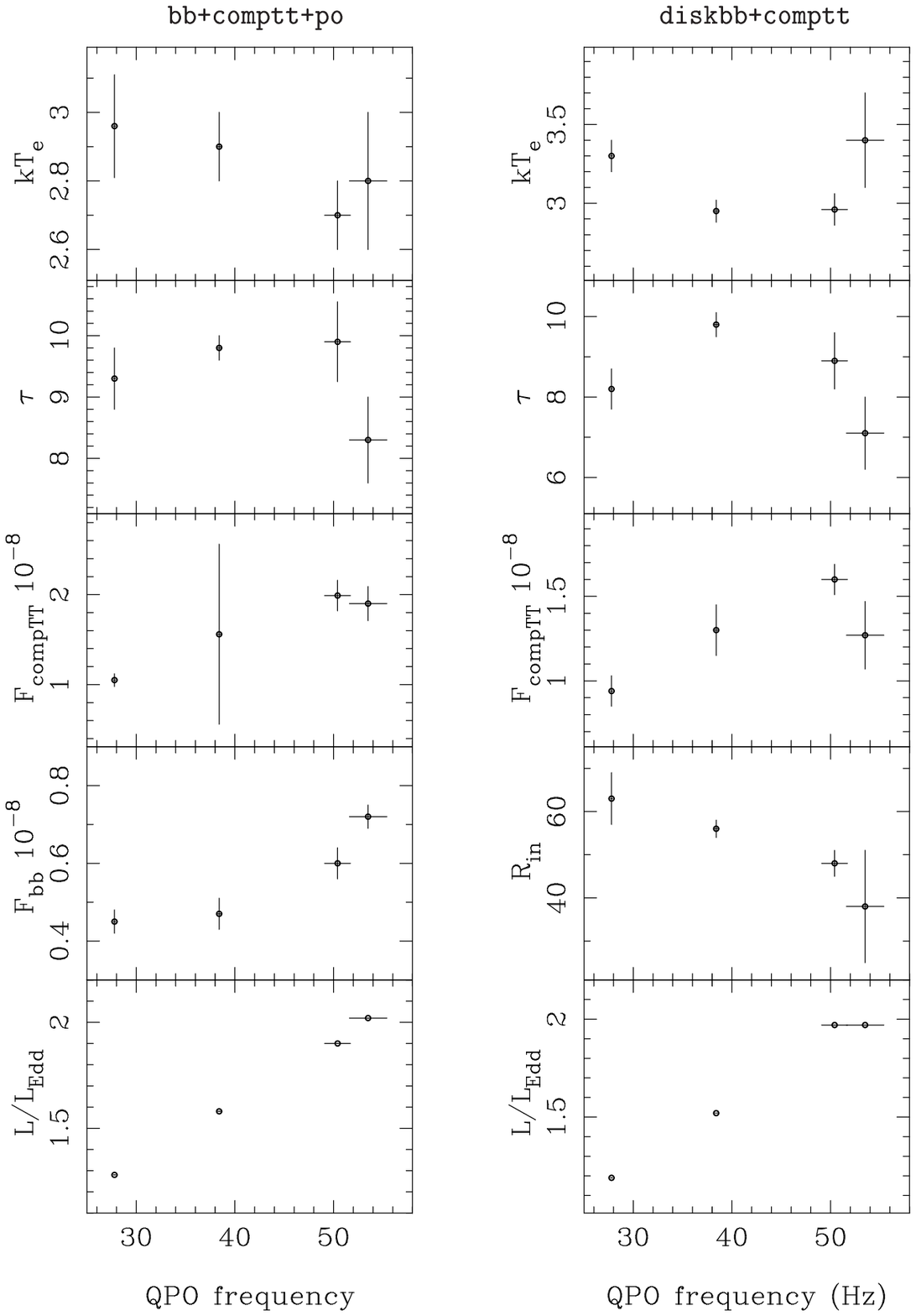} \caption{Left panels: best
fit parameters from the energy spectrum analysis with {\tt
{bb+comptt+po}}model versus the HBO frequency. Right panels:
best fit parameters from the energy spectrum analysis with {\tt
{diskbb+comptt}}model versus the HBO frequency.} \label{fig8}
\end{figure}

\begin{table}[ht!]
\begin{tiny}
\begin{center}
\label{tab1}
\begin{tabular}{c|ccc|cc|ccc|cc}
\tableline
 Z & \multicolumn{3}{c|}{LFN}
   & \multicolumn{2}{c|}{HFN}
   & \multicolumn{3}{c|}{HBO}
   & $\chi^2$ & DOF \\

   & $\alpha$ & $\nu$ & rms
   & $\nu$ & rms
   & $\nu$ & $\Delta\nu$ & rms
   & & \\

   & & Hz & $\%$
   & Hz & $\%$ & Hz
   & Hz & $\%$
   & & \\

\tableline \tableline

hb1 & $0.068 \pm 0.009$& $4.0 \pm 0.2$
    & $4.17 \pm 0.04$ & $32 \pm 2$ & $3.8 \pm 0.4$
    & $27.8 \pm 0.1$ & $6.7 \pm 0.3$ & $2.79 \pm 0.07$
    & $270$ & $65$ \\

hb2 & $0.264 \pm 0.004$& $6.8 \pm 0.3$
    & $4.97 \pm 0.04$ & $34 \pm 3$ & $3.6 \pm 0.4$
    & $38.4 \pm 0.2$ & $7.8 \pm 0.8$ & $2.00 \pm 0.09$
    & $330$ & $65$ \\

hb3 & $0.299 \pm 0.006$& $13.7 \pm 1.1$
    & $4.67 \pm 0.12$ & $41 \pm 10$ & $2.0 \pm 1.6$
    & $48.0 \pm 0.5$ & $16.8 \pm 2.3$ & $1.9 \pm 0.1$
    & $470$ &$65$ \\

hb4 & $0.326 \pm 0.008$& $20.0 \pm 1.1$
    & $4.62 \pm 0.06$ & $57 \pm 141$ & $0.9 \pm 0.9$
    & $50.4 \pm 1.3$ & $16.5 \pm 3.5$ & $1.7 \pm 0.2$
    & $222$ & $65$ \\

\tableline

hb1$_h$ & $-0.023 \pm 0.012$& $3.6 \pm 0.1$
       & $6.54 \pm 0.08$ & $28 \pm 2$ & $5.6 \pm 0.7$
       & $27.6 \pm 0.1$ & $6.7 \pm 0.4$ & $4.3 \pm 0.1$
       & $208$ & $65$ \\

hb2$_h$ & $0.194 \pm 0.007$ & $6.2 \pm 0.3$
       & $7.25 \pm 0.07$ & $33 \pm 3$ & $5.7 \pm 0.7$
       & $38.2 \pm 0.3$ & $7.0 \pm 1.1$ & $2.9 \pm 0.2$
       & $209$ & $65$ \\

hb3$_h$ & $0.243 \pm 0.007$ & $13.8 \pm 0.7$
       & $6.74 \pm 0.13$ & $41 \pm 11$ & $2.6 \pm 1.5$
       & $47.6 \pm 0.7$ & $19.6 \pm 2.7$ & $3.1 \pm 0.2$
       & $295$ &$65$ \\

hb4$_h$ & $0.263 \pm 0.018$ & $17.4 \pm 2.3$
       & $6.34 \pm 0.30$ & $53 \pm 70$ & $2.2 \pm 4.0$
       & $50.1 \pm 1.8$ & $15 \pm 6$ & $2.4 \pm 0.5$
       & $175$ & $65$ \\

\tableline
\end{tabular}
\caption{Results of the HB power spectral fits of the PCA
data. The power spectra hb1, hb2, hb3, hb4 were extracted
using data from all energy range, the power spectra
hb1$_h$,hb2$_h$, hb3$_h$, hb4$_h$ were extracted using
photons with energies above 5.8~keV.}
\end{center}
\end{tiny}
\end{table}

\begin{table}[ht!]
\begin{tiny}
\begin{center}
\vskip 0.5cm \label{tab2}
\begin{tabular}{c|cc|cc|ccc|ccc|cc}
\tableline
 Z & \multicolumn{2}{c|}{VLFN}
     & \multicolumn{2}{c|}{HFN}
     & \multicolumn{3}{c|}{HBO}
     & \multicolumn{3}{c|}{NBO} & $\chi^2$ & $DOF$ \\

     & $\alpha$ & rms & $\nu$
     & rms & $\nu$ & $\Delta\nu$
     & rms & $\nu$ & $\Delta\nu$
     & rms & & \\

     & & $\%$ & Hz
     & $\%$ & Hz & Hz
     & $\%$ & Hz & Hz
     & $\%$ & & \\

\tableline \tableline

nb1 & $2.3 \pm 0.1$ & $10.5 \pm 0.5$ & $11 \pm 2$
     & $2\pm 0.3$ & $53.5 \pm 1.9$ & $23 \pm 8$
     & $1.8 \pm 0.2$ & $5.0 \pm 0.2$ & $2.7 \pm 0.8$
     & $1.2 \pm 0.2$ & $41$ & $40$ \\

nbfb & $1.98 \pm 0.02$ & $9.6 \pm 0.1$ & $8.9 \pm 1.3$
     & $1.2 \pm 0.1$ & &
     & & &
     & & $32$ & $17$ \\

fb1 & $2.04 \pm 0.02$ & $17.4 \pm 0.1$ & $7.6 \pm 3.2$
     & $1.1 \pm 0.4$ & &
     & & &
     & & $34$ & $17$ \\

\tableline

nb1$_h$  & $1.92 \pm 0.23$ & $6.7 \pm 0.7 $ & $18 \pm 7$
         & $2.3 \pm 1.9$ & $56.5 \pm 5.4$ & $32 \pm 16$
         & $3.1 \pm 0.6$ & $5.1 \pm 0.2$ & $3.5 \pm 0.9$
         & $2.6 \pm 0.4$ & $34$ & $40$ \\

nbfb$_h$ & $2.06 \pm 0.06$ & $13.4 \pm 0.2 $ & $12.0 \pm 2.9$
         & $1.9 \pm 0.4 $ & &
         & & &
         & & $20$ & $17$ \\

fb1$_h$  & $2.18 \pm 0.07$ & $25.6\pm 0.6$ & $7.4 \pm 5.0$
         & $1.7\pm 0.9$ & &
         & & &
         & & $12$ & $17$ \\

\tableline
\end{tabular}
\caption{Results of the NB and FB power spectral fits of the PCA data}
\end{center}
\end{tiny}
\end{table}

\begin{table}[ht!]
\tiny
\begin{center}
\begin{tabular}{l|c|c|c|c|c|c}
\tableline \tableline Spectrum & E & E ($N_{\rm H}$ fixed) &F &
C & B & A \\

 Z region & $\subset$ {\tt hb1} & $\subset$ {\tt hb1} & $\subset$ {\tt hb2}
 & $\subset$ {\tt hb4} & $\subset$ {\tt nb1} & $\subset$ {\tt
nbfb} \\ \tableline

$N_{\rm H}$ $\rm (\times 10^{21}\;cm^{-2})$
 & $3.9 \pm 0.4$ & $2.0 (fixed)$
 & $1.9^{+0.9}_{-0.5}$ & $1.5^{+2}_{-0.6}$
 & $1.9^{+1.7}_{-0.5}$ & $2.2^{+1.7}_{-0.5}$ \\

$kT_{\rm BB}$ (keV)
 & $0.53 \pm 0.02$ & $0.50 \pm 0.01$
 & $0.49 \pm 0.03$ & $0.52 \pm 0.02$
 & $0.55 \pm 0.02$ & $0.51 \pm 0.03$ \\

$R_{\rm BB}$ (km)
 & $56 \pm 6$ & $90 \pm 9$
 & $70 \pm 9$ & $70 \pm 6$
 & $69 \pm 5$ & $76 \pm 9$ \\

$k T_{\rm W}$ (keV)
 & $1.06 \pm 0.04$ & $1.01 \pm 0.03$
 & $0.94 \pm 0.04$ & $1.05 \pm 0.05$
 & $1.08 \pm 0.03$ & $0.99 \pm 0.04$ \\

$k T_{\rm e}$ (keV)
 & $3.0 \pm 0.2$ & $3.0 \pm 0.2$
 & $2.9 \pm 0.1$ & $2.7 \pm 0.1$
 & $2.8 \pm 0.2$ & $2.4^{+0.4}_{-0.2}$ \\

$\tau$
 & $9.1 \pm 0.6$ & $9.3 \pm 0.5$
 & $9.8^{+0.3}_{-0.1}$ & $9.9^{+0.8}_{-0.5}$
 & $8.3 \pm 0.7$ & $9.1^{+0.6}_{-1.3}$ \\

$N_c$
 & $0.64 \pm 0.05$ & $0.70 \pm 0.05$
 & $1.11 ^{+0.9}_{-0.5}$ & $1.53 \pm 0.13$
 & $1.47 \pm 0.15 $ & $1.62 \pm 0.3 $ \\

%$R_{\rm W}$ (km) d=8.7 kpc & $25\pm $ & & $26 \pm $ &
 %$23\pm $ & $25 \pm $ & $28 \pm 4 $\\

$R_{\rm W}$ (km) (d=11.6 kpc)
 & $33 \pm 3 $ & $27 \pm 3 $
 & $35 \pm 4 $ & $30 \pm 4 $
 & $33 \pm 3 $ & $38 \pm 5 $ \\

$y$
 & $1.94 \pm 0.28 $ & $2 \pm 0.30 $
 & $2.18 \pm 0.12 $ & $2.07 \pm 0.30 $
 & $1.51 \pm 0.28 $ & $1.56 \pm 0.36 $ \\

Photon Index $\alpha$
 & $3.2 \pm 0.2 $ & $2.76 \pm 0.1 $
 & $2.8^{+0.3}_{-0.2}$ & $2.5\pm 0.3$
 & $2.7^{+0.3}_{-0.2}$ & $2.8 \pm 0.3$ \\

Power-law $N$
 & $2.4 \pm 0.5$ & $0.58 \pm 0.6$
 & $0.4^{+0.5}_{-0.2}$ & $0.4^{+1.3}_{-0.3}$
 & $0.5^{+1.6}_{-0.2}$ & $0.7^{+1.3}_{-0.4}$ \\

$E_{\rm line}$ (keV)
 & $1.06 \pm0.02$ & $0.97^{+0.05}_{-0.03}$
 & $0.86 \pm 0.16$ & $1.02^{+0.08}_{-0.12}$
 & $0.97^{+0.06}_{-0.09}$ & $0.94^{+0.08}_{-0.06}$ \\

$\sigma$ (keV)
 & $0.09 \pm 0.05$ & $0.23 \pm 0.04$
 & $0.33 \pm 0.07$ & $0.2^{+0.09}_{-0.05}$
 & $0.27^{+0.04}_{-0.14}$ & $0.24 \pm 0.05$ \\

$I_{\rm line}$ (ph cm$^{-2}$ s$^{-1}$)
 & $0.13\pm0.04$ & $0.34 \pm 0.08$
 & $0.6^{+0.4}_{-0.3}$ & $ 0.23^{+0.25}_{-0.08}$
 & $0.44 \pm 0.13$ & $0.5 \pm 0.3$ \\

line $Eq. W.$ (eV)
 & 44 & 176
 & 242 & 115
 & 186 & 181 \\

$E_{\rm Fe}$ (keV)
 & $6.80 \pm 0.1$ & $6.80 \pm 0.1$
 & $6.5^{+0.25}_{-0.3}$ & $6.7 \pm 0.1$
 & $6.69 (fixed)$ & $6.74 (fixed)$ \\

$\sigma$ (keV)
 & $0.25 \pm 0.1$ & $0.24 \pm 0.14$
 & $1.0^{+0.6}_{-0.3} $ & $0. (fix)$
 & $0. (fix)$ & $0. (fix)$ \\

$I_{\rm Fe}$ (ph cm$^{-2}$ s$^{-1}$)

 & $(3.7\pm 1) \times 10^{-3}$ & $(3.6 \pm 1 ) \times 10^{-3}$
 & $(15 \pm 5 ) \times 10^{-3}$ & $1.5 \times 10^{-3} (fix)$
 & $1.2 \times 10^{-3} (fix)$ & $1.6 \times 10^{-3} (fix)$ \\

Fe $Eq. W.$ (eV)

 & 44 & 44
 & 115 & 10
 & 8 & 13 \\

%Black body Flux

$F_{\rm bb}$ ($\times 10^{-8}$ ergs cm$^{-2}$ s$^{-1}$)
 & $(0.39 \pm 0.03)$ & $(0.45 \pm 0.03)$
 & $(0.47 \pm 0.04)$ & $(0.60 \pm 0.04)$
 & $(0.72 \pm 0.3) $ & $(0.63 \pm 0.05)$ \\

%Flux
$F_{\rm tot ab}$ ($\times 10^{-8}$ergs cm$^{-2}$ s$^{-1}$)
 & $1.53$ & $1.53$
 & $2.00$ & $2.60$
 & $2.62$ & $2.29$ \\

%Flux
$F_{\rm tot}$ ($\times 10^{-8}$ergs cm$^{-2}$ s$^{-1}$)
 & $3.58$ & $1.94$
 & $2.39$ & $2.87$
 & $3.06$ & $2.83$ \\

% % Black body Flux
$F_{\rm bb} /F_{\rm tot} (\%)$
 & $11$ & $23$
 & $20$ & $21$
 & $23$ & $22$ \\

%Luminosity % %$L_{\rm tot}$ ( d=8.7 kpc)
 % & 1.33$L_{edd}$ &
  %& 0.89$L_{edd}$ & 1.07$L_{edd}$
  %& 1.14$L_{edd}$ & 1.05$L_{edd}$ \\

%Luminosity

$L_{\rm tot}$ ( d=11.6 kpc)
 & 2.37$L_{edd}$ & 1.28$L_{edd}$
 & 1.58$L_{edd}$ & 1.90$L_{edd}$
 & 2.02$L_{edd}$ & 1.87$L_{edd}$ \\

$\chi^2_{\rm red}$ (d.o.f.)
 & 1.68 (113) & 1.80 (114)
 & 1.19 (119) & 1.18 (117)
 & 1.09 (116) & 1.00 (115) \\

$F$-{\rm test 1} (fe)
 & $6.1 \times 10^{-17}$ & $5.4 \times 10^{-13}$
 & $5.4 \times 10^{-13}$ & $2.3 \times 10^{-2}$
 & $1.6 \times 10^{-2}$ & $8.3 \times 10^{-3}$ \\

$F$-{\rm test 2} (po)
 & $7.4 \times 10^{-16}$ & $5.1 \times 10^{-15}$
 & $8.5 \times 10^{-6}$ & $2.7 \times 10^{-5}$
 & $6.3 \times 10^{-10}$ & $1.1 \times 10^{-14}$ \\
\tableline
\end{tabular}

\caption{ Results of the fit of Cyg~X-2 spectra in the energy
band 0.2--80~keV. The model consists of blackbody, {\tt comptt},
power law and two Gaussian emission line. For each spectrum the
corresponding PCA HC-I region is indicated. Uncertainties are at
the 90\% confidence level for a single parameter. The powerlaw
normalization is in units of ph keV$^{-1}$ cm$^{-2}$ s$^{-1}$ at
1~keV. The total flux is in the 0.2--200~keV energy range.
F-test 1 and F-test 2 are the probabilities of chance
improvement when an iron line and power law are included in the
spectral model. The relativistically corrected Eddington
Luminosity for a 1.9 $M_{\odot}$ neutron star with a peak
photospheric surface radius of 26 km and material with cosmic
abundances is 2.43$\times 10^{38}$ ergs s$^{-1}$ (Lewin et~al.\
1993).} \label{tab3}

\end{center}
\end{table}

\newpage

\begin{table}[ht!]
\tiny
\begin{center}
 \vskip 0.5cm

\label{tab4}
\begin{tabular}{l|c|c|c|c|c}
\tableline \tableline
 Spectrum & E & F & C & B & A \\
  Z region & $\subset$ {\tt hb1} & $\subset$ {\tt hb2} & $\subset$ {\tt hb4} & $\subset$ {\tt nb1} & $\subset$ {\tt
nbfb} \\ \tableline

$N_{\rm H}$ $\rm (\times 10^{21}\;cm^{-2})$
 & $1.88 \pm0.05$ & $1.96 \pm 0.04$
 & $2.2 \pm 0.1$ & $2.0 \pm 0.1$
 & $2.1\pm 0.1$ \\

$k T_{\rm in}$ (keV)
 & $0.79 \pm 0.03$ & $0.87 \pm 0.01$
 & $1.02 \pm 0.02$ & $1.2 \pm 0.3$
 & $1.07 \pm 0.01$ \\

$K=(R_{in}/D)^2 cos i$ (D=d/10 kpc)
 & $1016^{+150}_{-200}$ & $800\pm40$
 & $595^{+50}_{-80}$ & $370^{+300}_{-190}$
 & $560 \pm 30$ \\

$R_{\rm in}$ (km) (d=11.6 kpc, $i=70^\circ$)
 & $63 \pm 6$ & $56 \pm 2$
 & $48 \pm 3$ & $38 \pm 13$
 & $47 \pm 2$ \\

$k T_{\rm W}$ (keV)
 & $1.15 \pm 0.04$ & $1.09 \pm 0.02$
 & $1.26 \pm 0.02$ & $1.36 \pm 0.1$
 & $1.25 \pm 0.04$ \\

$k T_{\rm e}$ (keV)
 & $3.3 \pm 0.1$ & $2.95^{+0.05}_{-0.09}$
 & $2.96 \pm 0.10$ & $3.3 \pm 0.3$
 & $3.1^{+0.8}_{-0.4}$ \\

$\tau$
 & $8.2 \pm 0.5$ & $9.8 \pm 0.3$
 & $8.9 \pm 0.7$ & $7.1 \pm 0.9$
 & $7.0 \pm 1.5$ \\

$N_c$
 & $0.53 \pm 0.05$ & $0.84^{+0.4}_{-0.2}$
 & $1.01 \pm 0.06$ & $0.7^{+0.3}_{-0.1}$
 & $0.7 \pm 0.15$ \\

$y$
 & $1.74 \pm 0.22 $ & $2.22 \pm 0.15 $
 & $1.83 \pm 0.30 $ & $1.30 \pm 0.35 $
 & $1.19 \pm 0.40 $ \\

$R_{\rm W}$ (km) d=11.6 kpc
 & $21 \pm 3 $ & $25 \pm 4 $
 & $23 \pm 4 $ & $21 \pm 3 $
 & $25 \pm 4 $ \\

$E_{\rm line}$ (keV)
 & $1.08 \pm 0.02$ & $1.96 \pm 0.04$
 & $1.06 \pm 0.07$ & $1.10 \pm 0.04$
 & $1.03 \pm 0.04$ \\

$\sigma$ (keV)
 & $0.09 \pm 0.05$ & $0.33 \pm 0.07$
 & $0.2^{+0.09}_{-0.05}$ & $0.13\pm 0.05$
 & $0.11 \pm 0.05$ \\

I$_{\rm line}$ (ph cm$^{-2}$ s$^{-1}$)
 & $0.06 \pm 0.04$ & $0.08^{+0.04}_{-0.06}$
 & $0.05^{+0.08}_{-0.05}$& $0.44 \pm 0.13$
 & $0. \pm 0.3$ \\

line Eq. W. (eV)
 & 33 & 28
 & 19 & 37
 & 35 \\

$E_{\rm Fe}$ (keV)
 & $6.8 \pm 0.1$ & $6.4^{+0.1}_{-0.2}$
 & & $6.66 \pm 0.16$
 & $6.7^{+0.2}_{-0.5}$ \\

$\sigma$ (keV)
 & $0.27 \pm 0.1$ & $1.13 \pm 0.15$
 & & $0.(fix)$
 & $0. (fix)$ \\

$I_{\rm Fe}$ (ph cm$^{-2}$ s$^{-1}$)
 & $(4.0 \pm 0.8) \times 10^{-3}$ & $(19 \pm 2 ) \times 10^{-3}$
 & & $(1.3^{+4}_{-0.8} )\times 10^{-3}$
 & $(2.2^{+8.2}_{-1.2}) \times 10^{-3} $\\

Fe $Eq. W.$ (eV)
 & 48 & 140
 & & 8
 & 17 \\

$F_{\rm tot ab}$ ($\times 10^{-8}$ ergs cm$^{-2}$ s$^{-1}$)
 & $1.53$ & $2.00$
 & $2.60$ & $2.62$
 & $2.29$ \\

%Flux
$F_{\rm tot}$ ($\times 10^{-8}$ ergs cm$^{-2}$ s$^{-1}$)
 & $1.80$ & $2.30$
 & $2.97$ & $2.98$
 & $2.68$ \\

%Luminosity
$L_{\rm tot}$ (d=11.6 kpc)
 & 1.19$L_{edd}$ & 1.52$L_{edd}$
 & 1.97$L_{edd}$ & 1.97$L_{edd}$
 & 1.77$L_{edd}$ \\

$\chi^2_{\rm red}$ (d.o.f.)
 & 1.72 (112) & 1.16 (118)
 & 1.34 (116) & 1.19 (112)
 & 1.25 (114) \\
\tableline
\end{tabular}

\caption{Results of the fit of Cyg~X-2 spectra in the energy
band 0.2--80~keV. The model consists of multicolor disk, {\tt
comptt}, and two Gaussian emission line. For each spectrum the
corresponding PCA HC-I region is indicated. Uncertainties are at
the 90\% confidence level for a single parameter. The total flux
is in the 0.2--100~keV energy range.}

\end{center}
\end{table}

\end{document}